\documentclass[twocolumn,aps,superscriptaddress,showpacs,floatfix]{revtex4}
\usepackage{amssymb}
\usepackage{amsmath}
\usepackage{graphicx}
\usepackage[normalem]{ulem}
\usepackage[dvips]{color}
\usepackage{bm}
\usepackage{longtable}

\renewcommand\sout{\bgroup \color{red} \ULdepth=-.5ex \ULset}

\begin{document}

\title{Extracting the nuclear symmetry potential and energy from neutron-nucleus scattering data}

\author{Xiao-Hua Li}
\affiliation{INPAC, Department of Physics and Shanghai Key Laboratory for Particle
Physics and Cosmology, Shanghai Jiao Tong University, Shanghai 200240, China}
\affiliation{School of Nuclear Science and Technology, University of
South China, Hengyang, Hunan 421001, China }
\author{Bao-Jun Cai}
\affiliation{INPAC, Department of Physics and Shanghai Key Laboratory for Particle
Physics and Cosmology, Shanghai Jiao Tong University, Shanghai 200240, China}
\author{Lie-Wen Chen\footnote{%
Corresponding author: lwchen$@$sjtu.edu.cn}}
\affiliation{INPAC, Department of Physics and Shanghai Key Laboratory for Particle
Physics and Cosmology, Shanghai Jiao Tong University, Shanghai 200240, China}
\affiliation{Center of Theoretical Nuclear Physics, National Laboratory of Heavy Ion
Accelerator, Lanzhou 730000, China}
\author{Rong Chen}
\affiliation{INPAC, Department of Physics and Shanghai Key Laboratory for Particle
Physics and Cosmology, Shanghai Jiao Tong University, Shanghai 200240, China}
\affiliation{Department of Physics, Arizona State University, Arizona 85287-1504, USA}
\author{Bao-An Li}
\affiliation{Department of Physics and Astronomy, Texas A$\&$M University-Commerce,
Commerce, Texas 75429-3011, USA}
\affiliation{Department of Applied Physics, Xi'an Jiaotong University, Xi¡¯an 710049, China}
\author{Chang Xu}
\affiliation{Department of Physics, Nanjing University, Nanjing 210008, China}
\date{\today}

\begin{abstract}
According to the Hugenholtz-Van Hove theorem, the symmetry energy and its density slope
can be decomposed uniquely in terms of the single-nucleon potential in asymmetric
nuclear matter which, at normal density, can be constrained by the nucleon optical
model potential extracted from analyzing the nucleon-nucleus scattering data. To more
accurately extract information on the symmetry energy and its density slope at normal
density from neutron-nucleus scattering data, going beyond the well-known Lane potential,
we include consistently the second order terms in isospin asymmetry in both the optical
model potential and the symmetry energy decomposition. We find that the strength of the
second-order symmetry potential $U_\mathrm{sym,2}$ in asymmetric nuclear matter is
significant compared to the first-order one $U_\mathrm{sym,1}$, especially at high
nucleon momentum. While the $U_\mathrm{sym,1}$ at normal density decreases with nucleon
momentum, the $U_\mathrm{sym,2}$ is found to have the opposite behavior. Moreover, we
discuss effects of the $U_\mathrm{sym,1}$ and $U_\mathrm{sym,2}$ on determining the
density dependence of the symmetry energy, and we find that the available
neutron-nucleus scattering data favor a softer density dependence of the symmetry energy
at normal density with a negligible contribution from the $U_\mathrm{sym,2}$.
\end{abstract}

\maketitle

\section{Introduction}
Knowledge about the density dependence of nuclear symmetry energy $E_\mathrm{sym}(\rho)$
is critical for understanding not only the structure and reaction of radioactive
nuclei, but also many interesting issues in astrophysics, such as the structure of
neutron stars and the mechanism of supernova explosions. Although significant
progress has been made in recent years in constraining the $E_\mathrm{sym}(\rho)$,
large uncertainties still exist, even around the saturation density of nuclear matter
(See, e.g., Refs.~\cite{LiBA98,Bar05,Ste05,ChenLW07,LCK08,Tsa12,Lat12,ChenLW12,LiBA12}).
It is thus of critical importance to better understand and reduce the uncertainties in
constraints placed on the $E_\mathrm{sym}(\rho)$ based on various model analyses of
experimental data.

Recently, based on the Hugenholtz-Van Hove (HVH) theorem~\cite{Hug58},
it has been shown that both the magnitude of $E_\mathrm{sym}(\rho)$
and its density slope $L(\rho)$ can be analytically decomposed in terms
of the single-nucleon potential in asymmetric nuclear matter~\cite{XuC10,XuC11,Che12}.
Moreover, the more general decomposition of $E_\mathrm{sym}(\rho)$ and
$L(\rho)$ in terms of the Lorentz covariant nucleon self-energies within the
relativistic covariant framework has also been obtained~\cite{Cai12}.
These decompositions help us better understand the underlying isospin
dependence of strong interaction contributing to the symmetry energy. They also
provide a new and physically transparent approach to extract information about
the symmetry energy from certain types of experimental data. In particular, at
saturation density $\rho_0$, the single-nucleon potentials or nucleon self-energies
in asymmetric nuclear matter can be constrained by the nucleon optical model
potential (OMP) extracted from analyzing the nucleon-nucleus scattering data.
In Ref.~\cite{XuC10}, indeed, the constraints $E_\mathrm{sym}(\rho_0)=31.3\pm 4.5$ MeV
and $L(\rho_0)=52.7\pm 22.5$ MeV have been obtained simultaneously by
using the phenomenological isovector nucleon OMP from averaging the world
data from nucleon-nucleus scatterings, (p,n) charge-exchange reactions, and
single-particle energy levels of bound states available in the literature
since 1969.

In the usual optical model analyses, however, up to now only the first-order
symmetry (isovector) potential, i.e., the Lane potential~\cite{Lan62},
has been considered. On the other hand, using three well-established models
it has been shown recently that the contribution of the second-order symmetry
potential to $L(\rho_0)$ depends on the interactions used and generally it cannot
be simply neglected~\cite{Che12}. Unfortunately, to our best knowledge, so far
there is no empirical or experimental information on the second-order symmetry
potential. It is thus of great importance to extract experimental information
about the second-order symmetry potential and examine its effects on the density
dependence of the symmetry energy. Here we report results of the first work in
this direction.

\section{Theoretical Formulism }
\subsection{Global neutron optical model potential}

The optical model is fundamentally important for many aspects of
nuclear physics~\cite{Yon98}. It is the basis and starting point for many
nuclear model calculations and also is one of the most important
theoretical approaches in nuclear data evaluations and analyses.
Over the past years, a number of excellent local and global nucleon OMP
have been proposed~\cite{Kon03,Bec69,Var91,Wep09,Han10,LiX12}.
The phenomenological OMP for neutron-nucleus scattering $V(r,\mathcal{E})$
is usually defined as
\begin{align}\label{OMP1}
V(r,\mathcal{E})=&-V_\mathrm{v} f_\mathrm{r}(r)-i W_\mathrm{v}
f_\mathrm{v}(r)+i 4 a_\mathrm{s}W_\mathrm{s}
\frac{\mathrm{d}f_\mathrm{s}(r)}{\mathrm{d}r}\nonumber\\&+2\lambda\!\!\!{^-}_\pi^2
\frac{V_\mathrm{so}+iW_\mathrm{so}}{r}
\frac{\mathrm{d}f_\mathrm{so}(r)}{\mathrm{d}r}
\mathbf{S}\cdot\mathbf{L}\,,
\end{align}
where $V_\mathrm{v}$ and $V_\mathrm{so}$ are the depth of the real
part of the central and spin-orbit potential, respectively;
$W_\mathrm{v}$, $W_\mathrm{s}$ and $W_\mathrm{so}$ are the depth
of the imaginary part of the volume absorption, surface
absorption and spin-orbit potential, respectively; the
$f_{i}$ ($i=\mathrm{r,v,s,so}$) are the standard Wood-Saxon shape
form factors; $\mathcal{E}$ is the incident neutron energy in the
laboratory frame; $\lambda\!\!\!{^-}_\pi^2$ is the Compton wave length
of pion, and here we use $\lambda\!\!\!{^-}_\pi^2 = 2.0$ fm${^2}$.

In this work, to obtain information on the energy dependence of
the first-order symmetry potential $U_\mathrm{sym,1}$ and the
second-order symmetry potential $U_\mathrm{sym,2}$ in asymmetric nuclear
matter, we extend the traditional phenomenological OMP for neutron-nucleus
scattering to include the isospin dependent terms up to the second order
in $V_\mathrm{v}$, $W_\mathrm{s}$ and $W_\mathrm{v}$, and they are
parameterized as
\begin{align}
V_\mathrm{v}=&V_0+V_1\mathcal{E}
+V_2\mathcal{E}^2+(V_3+V_{3\mathrm{L}}\mathcal{E}) \frac{N-Z}{A}
\nonumber\\&+(V_4+V_{4\mathrm{L}}\mathcal{E})
\frac{(N-Z)^2}{A^2},\label{OMP2-1}\\
W_\mathrm{s}=&W_{\mathrm{s}0}+W_{\mathrm{s}1}\mathcal{E}+(W_{\mathrm{s}2}+W_{\mathrm{s}2\mathrm{L}}\mathcal{E})
\frac{N-Z}{A}
\nonumber\\&+(W_{\mathrm{s}3}+W_{\mathrm{s}3\mathrm{L}}\mathcal{E})
\frac{(N-Z)^2}{A^2},\label{OMP2-2}\\
 W_\mathrm{v}=&W_{\mathrm{v}0}+W_{\mathrm{v}1}\mathcal{E}
 +W_{\mathrm{v}2}\mathcal{E}^2+(W_{\mathrm{v3}}+W_{\mathrm{v3L}}\mathcal{E})\frac{N-Z}{A}
 \nonumber\\&+(W_{\mathrm{v4}}+W_{\mathrm{v4L}}\mathcal{E})
\frac{(N-Z)^2}{A^2},\label{OMP2-3}
\end{align}
where $-(V_0+V_1\mathcal{E}+V_2\mathcal{E}^2)\equiv\mathcal{U}_0(\mathcal{E})$
is the isospin-independent real part of the central potential while
$-(V_3+V_{3\mathrm{L}}\mathcal{E})\equiv\mathcal{U}_\mathrm{sym,1}(\mathcal{E})$ and
$-(V_4+V_{4\mathrm{L}}\mathcal{E})\equiv\mathcal{U}_\mathrm{sym,2}(\mathcal{E})$ are
the first- and second-order symmetry potentials of the real part of the central potential
in the optical model, respectively. The shape form factors $f_{i}$
is expressed as
\begin{align}\label{OMP3}
f_i(r)
=\left[1+\exp\left(\left.\left(r-r_i
\,A^{1/3}\right)\right/a_i\right)\right]^{-1},
\end{align}
with
\begin{align}\label{OMP4}
r_i=&r_{i0}+r_{i1} \,A^{-1/3},~~ a_i=a_{i0}+a_{i1} \,A^{1/3}.
\end{align}
In the above equations, $A = Z + N$ with $Z$ and $N$ being the number
of protons and neutrons of the target nucleus, respectively.

\subsection{Single-nucleon potential decomposition of $E_\mathrm{sym}(\rho)$ and $L(\rho)$}

The equation of state (EOS) of asymmetric nuclear matter
can be written as
\begin{eqnarray}\label{a1}
E(\rho,\delta)=E_0(\rho)+E_\mathrm{sym}(\rho)\delta^2+\mathcal{O}(\delta^4),
\label{EOS}
\end{eqnarray}
where $E_0(\rho)$ is the binding energy per nucleon in symmetric
nuclear matter; $\rho=\rho_{\mathrm{n}}+\rho_{\mathrm{p}}$ is the
baryon density and $\delta=(\rho_{\mathrm{n}}-\rho_{\mathrm{p}})/\rho$ is
the isospin asymmetry with $\rho_{\mathrm{n}}$ and $\rho_{\mathrm{p}}$
denoting the neutron and proton densities, respectively;
$E_\mathrm{sym}(\rho)$ is the nuclear symmetry energy and around
saturation density it can be expanded as
\begin{eqnarray}\label{a2}
E_\mathrm{sym}(\rho)=E_\mathrm{sym}(\rho_0)+L\chi+\mathcal{O}(\chi^2),~~\chi\equiv\frac{\rho-\rho_0}{3\rho_0},
\end{eqnarray}
with $L\equiv L(\rho_0)=\left.3\rho_0\frac{\partial
E_\mathrm{sym}(\rho)}{\partial
\rho}\right|_{\rho=\rho_0}$ being the density slope parameter.

In the non-relativistic framework, the single-nucleon energy
$\mathcal{E}(\rho,\delta,{\bf |k|})$ in asymmetric nuclear matter can be
expressed generally using the following dispersion relation
\begin{align}
\mathcal{E}_J(\rho,\delta,|{\bf {k}}|)=\frac{
|\mathbf{k}|^2}{2m}+U_J(\rho,\delta,|{\bf k}|,\mathcal{E}),~(J=\text{n or p})
\label{Dispersion}
\end{align}
where the single-particle potential $U_J(\rho,\delta,|{\bf k}|,\mathcal{E})$
can be expanded as a power series of $\delta$ as
\begin{align}\label{a5}
U_J(\rho,\delta,{\bf |k|},\mathcal{E})=&U_0(\rho,|{\bf
k}|,\mathcal{E})\notag\\
&+\sum_{i=1}U_{\mathrm{sym},i}(\rho,|{\bf
k}|,\mathcal{E})(\tau_3^J)^i\delta^i,
\end{align}
with the $i$th-order symmetry potential defined by
\begin{align}\label{a6}
&U_{\mathrm{sym},i}(\rho,|{\bf
k}|,\mathcal{E})=\frac{1}{i!}\frac{\mathrm{d}^i}{\mathrm{d}\delta^i}\left[\sum_J\frac{(\tau_3^J)^iU_J(\rho,\delta,|{\bf
k}|,\mathcal{E})}{2}\right]_{\delta=0},
\end{align}
and $U_0$ being the single-nucleon potential in symmetric nuclear matter.
In the above expressions, $\tau_3^J$ is the third component of the isospin
index and here we use the convention $\tau_3^{\mathrm{n}}=+1$ and $\tau_3^{\mathrm{p}}=-1$.
The nucleon mass $m$ is set to be $939$ MeV in the following calculations.

From the HVH theorem~\cite{Hug58}, the Fermi energy $\mathcal{E}^J_\mathrm{F}$
in asymmetric nuclear matter can be expressed as
\begin{align}
\mathcal{E}^J_{\mathrm{F}}(\rho,\delta,k^J_{\mathrm{F}})=\frac{\partial(
\rho E)}{\partial \rho_{J}},
\label{EFermi}
\end{align}
where $k^J_{\mathrm{F}}$ is the nucleon Fermi momentum expressed as
\begin{align}\label{a8}
k^J_\mathrm{F}=k_{\mathrm{F}}(1+\tau_3^J\delta)^{1/3},~~
k_{\mathrm{F}}=\left(\frac{3\pi^2\rho}{2}\right)^{1/3}.
\end{align}
Using Eqs.~\eqref{EOS}, \eqref{Dispersion} and \eqref{EFermi}, one can
decompose the symmetry energy in terms of single-nucleon potentials
as~\cite{XuC10,XuC11,Che12}
\begin{align}
E_{\mathrm{sym}}(\rho)=E^1_{\mathrm{sym}}(\rho )+E^2_{\mathrm{sym}}(\rho ),
\label{a10}
\end{align}
with
\begin{align}
E^1_{\mathrm{sym}}(\rho )=&\left.\frac{k^2_{\mathrm{F}}}{6m}+\frac{k_{\mathrm{F}}}{6}\frac{\partial
U_0(\rho, |{\bf k}|,\mathcal{E})}{\partial |{\bf
k}|}\right|_{k_{\mathrm{F}}}\notag\\
&+\frac{k_{\mathrm{F}}}{6}\left.\frac{\partial
\mathcal{E}}{\partial |{\bf
k}|}\right|_{k_{\mathrm{F}}}\cdot\left.\frac{\partial U_0(\rho,
|{\bf k}|,\mathcal{E})}{\partial
\mathcal{E}}\right|_{k_{\mathrm{F}}},\\
E^2_{\mathrm{sym}}(\rho )=&\frac{1}{2}U_{\mathrm{sym,1}}(\rho,|{\bf k}|,\mathcal{E})|_{k_{\mathrm{F}}}.
\end{align}
In the above decomposition, $E^1_{\mathrm{sym}}(\rho )$ represents the kinetic
energy part (including the contribution from the isoscalar nucleon effective mass)
of the symmetry energy while $E^2_{\mathrm{sym}}(\rho )$ is the contribution of
the first-order symmetry potential $U_{\mathrm{sym,1}}$. Similarly, for the density
slope parameter $L$ at an arbitrary density $\rho$, one can obtain five terms
with different characteristics~\cite{XuC10,XuC11,Che12}, i.e.,
\begin{align}
L(\rho)=L_1(\rho)+L_2(\rho)+L_3(\rho)+L_4(\rho)+L_5(\rho),
\label{a11}
\end{align}
with
\begin{align}
L_1(\rho)=&\frac{k^2_{\mathrm{F}}}{3m}+\left.\frac{k_{\mathrm{F}}}{6}\frac{\partial
U_0(\rho, |{\bf k}|,\mathcal{E})}{\partial |{\bf
k}|}\right|_{k_{\mathrm{F}}}\notag\\
&+\left.\frac{k_{\mathrm{F}}}{6}\frac{\partial
\mathcal{E}}{\partial |{\bf k}|}\cdot\frac{\partial U_0(\rho,|{\bf
k}|,\mathcal{E})}{\partial
\mathcal{E}}\right|_{k_{\mathrm{F}}},\\
L_2(\rho)=&\frac{k^2_{\mathrm{F}}}{6}\frac{\partial ^2U_0(\rho, |{\bf
k}|,\mathcal{E})}{\partial |{\bf
k}|^2}|_{k_{\mathrm{F}}}+\frac{k^2_{\mathrm{F}}}{6}\Bigg\{\frac{\partial^2\mathcal{E}}{\partial|{\bf
k}|^2}\cdot\frac{\partial U_0(\rho,|{\bf
k}|,\mathcal{E})}{\partial \mathcal{E}}
 \notag\\&+\left(\frac{\partial \mathcal{E}}{\partial |{\bf
k}|}\right)^2\cdot\frac{\partial ^2U_0(\rho,|{\bf
k}|,\mathcal{E})}{\partial^2
\mathcal{E}}\Bigg\}_{k_{\mathrm{F}}},\\
L_3(\rho)=&\frac{3}{2}U_{\mathrm{sym,1}}(\rho,
|{\bf k}|,\mathcal{E})|_{k_{\mathrm{F}}},\\
L_4(\rho)=&k_{\mathrm{F}}\left[\frac{\partial
U_{\mathrm{sym,1}}(\rho,|{\bf k}|,\mathcal{E})}{\partial |{\bf
k}|}+\frac{\partial \mathcal{E}}{\partial |{\bf
k}|}\cdot\frac{\partial U_{\mathrm{sym,1}}(\rho, |{\bf k}|,
\mathcal{E})}{\partial
\mathcal{E}} \right]_{k_{\mathrm{F}}},\\
L_5(\rho)=&3U_{\mathrm{sym},2}(\rho,|{\bf k}|,\mathcal{E})|_{k_{\mathrm{F}}}.
\end{align}
The above expressions clearly show what physics ingredients determine the
density dependence of the symmetry energy. More specifically, the $L_{1}(\rho )$
represents the kinetic contribution (including the effect from the isoscalar
nucleon effective mass), the $L_{2}(\rho )$ is due to the momentum
dependence of the isoscalar nucleon effective mass, the $L_{3}(\rho )$ is
directly from the first-order symmetry potential $U_{\mathrm{sym,1}}$,
the $L_{4}(\rho )$ relates to the momentum dependence of the $U_{\mathrm{sym,1}}$,
and the $L_{5}(\rho )$ is from the second-order symmetry potential $U_{\mathrm{sym,2}}$.

\subsection{The nuclear optical model potential and the symmetry potential
in asymmetric nuclear matter}

In order to obtain the values of the symmetry energy $E_{\mathrm{sym}}(\rho)$
and the slope parameter $L(\rho)$ at saturation density $\rho_0$, one
needs information about the energy (momentum) dependence of the single-nucleon potential,
i.e., $U_0$, $U_{\mathrm{sym,1}}$, $U_{\mathrm{sym,2}}$ in asymmetric nuclear matter
at $\rho_0$. The $U_0$, $U_{\mathrm{sym,1}}$, and $U_{\mathrm{sym,2}}$ at $\rho_0$
can be obtained from the real part of the central potential in the optical model,
i.e., $\mathcal{U}_0(\mathcal{E})$, $\mathcal{U}_\mathrm{sym,1}(\mathcal{E})$,
and $\mathcal{U}_\mathrm{sym,2}(\mathcal{E})$. In the following, we briefly
describe the connection between $U_0$, $U_{\mathrm{sym,1}}$, $U_{\mathrm{sym,2}}$
in asymmetric nuclear matter and $\mathcal{U}_0(\mathcal{E})$,
$\mathcal{U}_\mathrm{sym,1}(\mathcal{E})$, and $\mathcal{U}_\mathrm{sym,2}(\mathcal{E})$
in the optical model.

Both the single-particle potential $U_\mathrm{n}$ of a neutron in asymmetric nuclear
matter and the real part of the central potential $\mathcal{U}_\mathrm{n}$ in neutron
OMP can be expanded as a power series of $\delta$ as
\begin{align}\label{b9}
\Gamma_\mathrm{n}=\Gamma_0\,+\Gamma_\mathrm{sym,1}\,\delta\,+\Gamma_\mathrm{sym,2}\,\delta^{2}+\cdots.\,\,
(\Gamma=U,\,\mathcal{U})
\end{align}
The neutron single-particle energy in asymmetric nuclear matter satisfies the
following dispersion relation
\begin{eqnarray}\label{b10}
\mathcal{E}_\mathrm{n}=T_\mathrm{n}+U_\mathrm{n}(T_\mathrm{n}),
\end{eqnarray}
where $\mathcal{E}_\mathrm{n}$, $T_\mathrm{n}$ and $U_\mathrm{n}(T_\mathrm{n})$
denote the single-neutron energy, the kinetic energy and the potential
energy, respectively. In symmetric nuclear matter, the single-neutron
energy is
\begin{eqnarray}\label{b11}
\mathcal{E}=T+U_0(T).
\end{eqnarray}
Following Ref.~\cite{Dab64}, by expanding Eq.~\eqref{b10} in the power
series of $\delta$ to the second-order and using Eq.~\eqref{b9}
and Eq.~\eqref{b11}, one can easily obtain
\begin{align}
U_0&=\mathcal{U}_0,~~ U_{\mathrm{sym,1}}=\frac{\mathcal{U}_{\mathrm{sym,1}}}{\mu},~~\notag\\
U_{\mathrm{sym,2}}&=\frac{\mathcal{U}_{\mathrm{sym,2}}}{\mu}+\frac{\zeta\mathcal{U}_{\mathrm{sym,1}}}{\mu^2}+\frac{\vartheta\mathcal{U}_{\mathrm{sym,1}}^2}{\mu^3},
\label{b12}
\end{align}
with
\begin{align}
\mu&=1-\frac{\partial\mathcal{U}_0}{\partial\mathcal{E}},~~
\zeta=\frac{\partial\mathcal{U}_{\mathrm{sym,1}}}{\partial\mathcal{E}},~~
\vartheta=\frac{\partial^2\mathcal{U}_{\mathrm{sym,1}}}{\partial\mathcal{E}^2}.
\label{b13}
\end{align}
The relations of $U_0=\mathcal{U}_0$ and $U_{\mathrm{sym,1}}=\frac{\mathcal{U}_{\mathrm{sym,1}}}{\mu}$ were
firstly derived in Ref.~\cite{Dab64}.

\section{ RESULTS AND DISCUSSIONS}

\begin{table}[!htb]
\caption{The parameters $V_i$ and $W_{\text{v}i}$ ($i=\mathrm{0,1,2,3,3L,4,4L}$)
as well as $W_{\text{s}i}$ ($i=\mathrm{0,1,2,3,3L}$) obtained directly using
all the data in the original database in Ref.~\cite{LiX12}.}
\begin{tabular}{cccc}
\hline
  parameter & value & parameter  & value\\
  \hline
$V_0(\mathrm{MeV})$&55.306&$W_{\mathrm{v}0}(\mathrm{MeV})$&$-$1.7064\\
$V_1$&$-$0.341&$W_{\mathrm{v}1}$&0.203038\\
$V_2(\mathrm{MeV}^{-1})$&4.43$\times$$10^{-4}$&$W_{\mathrm{v}2}(\mathrm{MeV}^{-1})$&$-$7.21$\times$$10^{-4}$\\
$V_3(\mathrm{MeV})$&$-$20.051&$W_{\mathrm{v}3}(\mathrm{MeV})$&$-$7.19$\times$$10^{-4}$\\
$V_{\mathrm{3L}}$&0.236&$W_{\mathrm{v3L}}$&0.209\\
$V_4(\mathrm{MeV})$&$-$9.431&$W_{\mathrm{v}4}(\mathrm{MeV})$&$-$6.12$\times$$10^{-5}$\\
$V_{\mathrm{4L}}$&3.088$\times$$10^{-5}$&$W_{\mathrm{v4L}}$&1.24$\times$$10^{-6}$\\
$W_{\mathrm{s}0}(\mathrm{MeV})$&12.178&$W_{\mathrm{s2L}}$&2.40$\times$$10^{-2}$\\
$W_{\mathrm{s}1}$&$-$0.2302&$W_{\mathrm{s}3}(\mathrm{MeV})$&$-$6.63$\times$$10^{-6}$\\
$W_{\mathrm{s}2}(\mathrm{MeV})$&$-$19.025&$W_{\mathrm{s3L}}$&1.04$\times$$10^{-5}$\\

\hline \hline
\end{tabular}
\label{Tab1}
\end{table}
The detailed method for obtaining the optimal OMP parameters can
be found, e.g., in Refs.~\cite{LiX12,LiX07,She02}. All of the relevant
experimental data used in this work are summarized in Ref.~\cite{LiX12}.
In particular, the original database for searching the global
neutron OMP parameters are taken from Table I in
Ref.~\cite{LiX12}. To see the magnitude of the isospin dependent
second order terms in $V_\mathrm{v}$, $W_\mathrm{s}$ and $W_\mathrm{v}$, we
firstly obtain the optimal OMP parameters directly using all the data in the
original database in Ref.~\cite{LiX12}, and the results for the parameters
$V_i$ and $W_{\text{v}i}$ ($i=\mathrm{0,1,2,3,3L,4,4L}$) as well as
$W_{\text{s}i}$ ($i=\mathrm{0,1,2,3,3L}$) are listed in Table~\ref{Tab1}. One
can see that while the isospin dependent second order terms in $V_\mathrm{v}$
are significant, they are negligibly small in $W_\mathrm{s}$ and $W_\mathrm{v}$.

In order to evaluate the error of the OMP parameters,
we use the Monte Carlo method, i.e., randomly sample $10\%$ of the
total experimental data in the original database to form a new searching
database, from the latter we then obtain a new set of the OMP parameters.
This procedure is repeated $1000$ times, and the average values of the OMP
parameters and the corresponding errors are then evaluated statistically.
In the Monte Carlo estimate, we have neglected the isospin dependent
second order terms in $W_\mathrm{s}$ and $W_\mathrm{v}$ since they have
essentially no effect on our results.
\begin{table}[!htb]
\caption{The average values and the corresponding variances (errors) for the
parameter $V_i$ ($i=\mathrm{0,1,2,3,3L,4,4L}$) obtained from $1000$
samples of OMP parameters sets. The $V_i$ values in Ref.~\cite{LiX12}
are also included for comparison.}
\begin{tabular}{cccc}
\hline
  parameter & average value & variance & Ref.~\cite{LiX12}\\
  \hline
$V_0(\mathrm{MeV})$&55.415&0.3316&54.983\\
$V_1$&$-$0.337&1.163$\times$$10^{-2}$&$-$0.328\\
$V_2(\mathrm{MeV}^{-1})$&3.93$\times$$10^{-4}$&1.741$\times$$10^{-4}$&3.10$\times$$10^{-4}$\\
$V_3(\mathrm{MeV})$&$-$20.107&1.6044&$-$18.495\\
$V_{\mathrm{3L}}$&0.236&6.298$\times$$10^{-2}$&0.219\\
$V_4(\mathrm{MeV})$&$-$9.368&1.6296&-\\
$V_{\mathrm{4L}}$&3.088$\times$$10^{-5}$&1.3938 $\times$$10^{-4}$&-\\

\hline \hline
\end{tabular}
\label{Tab2}
\end{table}

\begin{figure}
\centering
\includegraphics[width=8cm]{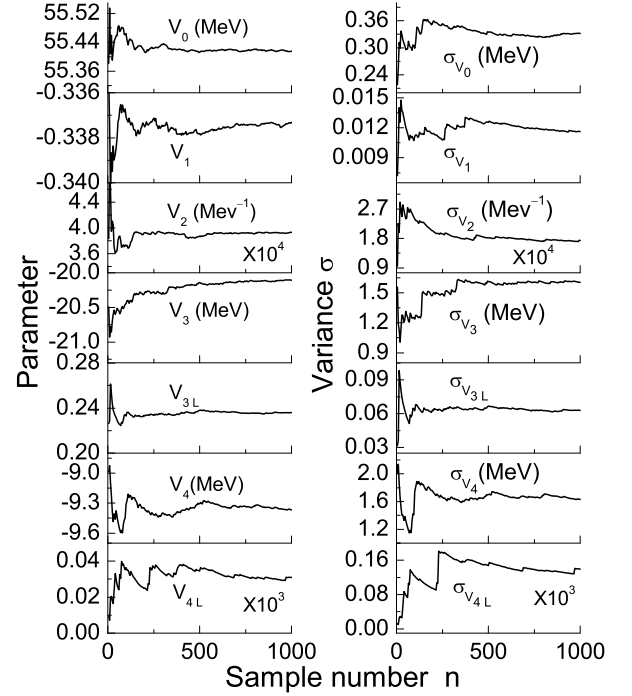}
\caption{The average values and variances (errors) of the
parameters $V_i(i=0,1,2,3,3\mathrm{L},4,4\mathrm{L})$ as a
function of the sample number $n$ of OMP parameters sets. }
\label{OptPara}
\end{figure}

Shown in Table~\ref{Tab2} are the average values and variances (errors) of
the parameters $V_i(i=\mathrm{0,1,2,3,3L,4,4L})$ obtained using the $1000$
samples of the OMP parameter sets. For comparison, we also include in
Table~\ref{Tab2} the $V_i$ values of Ref.~\cite{LiX12} where the higher-order
terms $V_i(i=\mathrm{4,4L})$ were neglected.  One can see that including the
higher-order terms $V_i(i=\mathrm{4,4L})$ enhances the magnitude of the
$V_i(i=\mathrm{3,3L})$ although the average values of $V_i(i=\mathrm{0,1,2,3,3L})$
are still consistent with the values obtained directly from all the data in 
the original database in Ref.~\cite{LiX12} without $V_i(i=\mathrm{4,4L})$.
To be more clear, the average values and variances of $V_i(i=\mathrm{0,1,2,3,3L,4,4L})$
are also shown in Fig.~\ref{OptPara} as a function of the sample number $n$ of
the OMP parameters sets. It is seen that both the average values and the variances
have all well converged at $n=1000$.

\begin{figure}
\centering
\includegraphics[width=8cm]{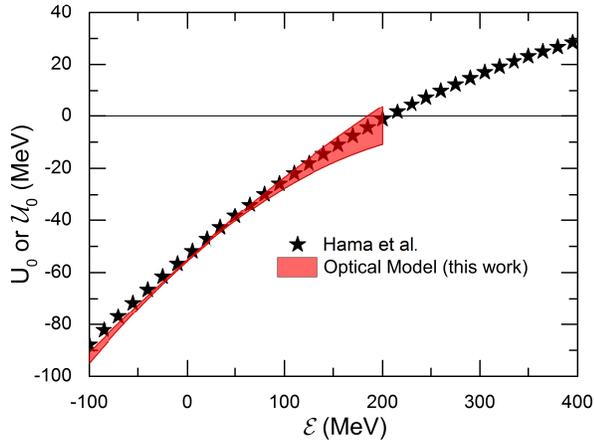}
\caption{(Color online) Energy dependence of single-nucleon
potential $U_0$ ($\mathcal{U}_0$) in symmetric nuclear matter obtained
from our OMP parameters. The results of the Schr$\ddot{\mathrm{o}}$dinger equivalent
potential obtained by Hama \textit{et al}~\cite{Ham90} from the nucleon-nucleus
scattering data are also included for comparison.}
\label{U0}
\end{figure}

\begin{figure}
\centering
\includegraphics[width=8cm]{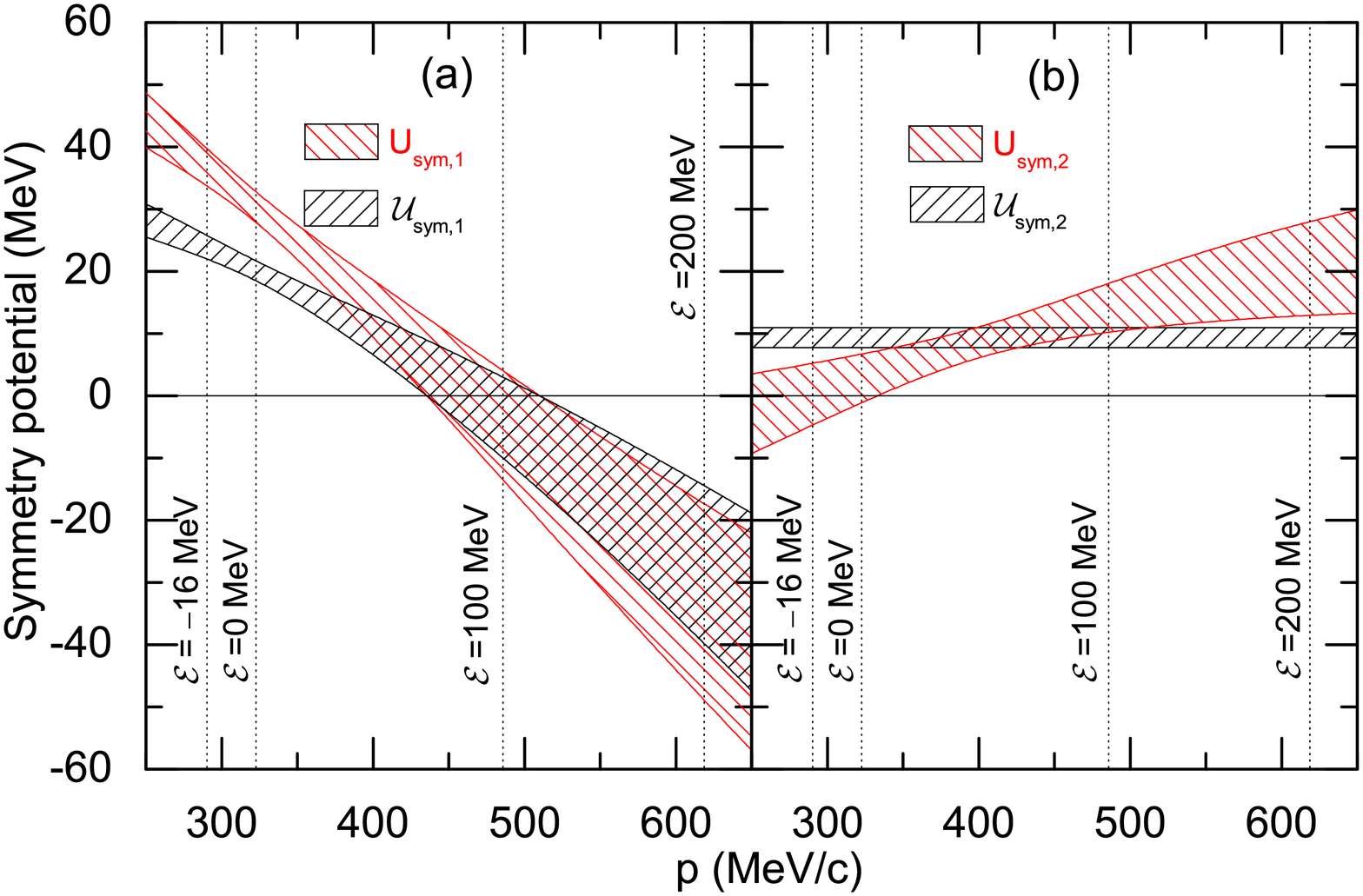}
\caption{(Color online) Momentum dependence of $\mathcal{U}_{\mathrm{sym,1}}$ ($U_\mathrm{sym,1}$) (a)
and $\mathcal{U}_{\mathrm{sym,2}}$
($U_\mathrm{sym,2}$) (b). The corresponding momenta at
$\mathcal{E}=-16$, $0$, $100$ and $200$ MeV are indicated by dotted lines.}
\label{Usym}
\end{figure}

Shown in Fig.~\ref{U0} is the energy dependence of the single-nucleon isoscalar
potential $U_0$ ($\mathcal{U}_0$) obtained from our OMP parameters. For comparison,
the results of the Schr$\ddot{\mathrm{o}}$dinger equivalent potential obtained by Hama
\textit{et al}~\cite{Ham90} from the nucleon-nucleus scattering data based on
the Dirac phenomenology are also shown. It is seen clearly that our isoscalar
potential is in good agreement with that from the Dirac phenomenology. We limit
the comparison up to 200 MeV since the experimental data of the neutron-nucleus
elastic scattering angular distributions used in the present work is in the beam
energy range of about $0-200$ MeV~\cite{LiX12}.

From the OMP parameters $V_i$($i=\mathrm{0,1,2,3,3L,4,4L}$), we can first
obtain the energy (momentum) dependence of the $\mathcal{U}_{\mathrm{sym,1}}$
and $\mathcal{U}_{\mathrm{sym,2}}$, and then the energy (momentum)
dependence of $U_\mathrm{sym,1}$ and $U_\mathrm{sym,2}$ using Eq.~\eqref{b12}.
Shown in Fig.~\ref{Usym} is the momentum dependence of $\mathcal{U}_{\mathrm{sym,1}}$,
$U_\mathrm{sym,1}$, $\mathcal{U}_{\mathrm{sym,2}}$, and $U_\mathrm{sym,2}$.
One can see that the $U_\mathrm{sym,1}$ is larger than the $\mathcal{U}_{\mathrm{sym,1}}$
at lower momenta (energies) while the $U_\mathrm{sym,1}$ becomes smaller than
the $\mathcal{U}_{\mathrm{sym,1}}$ at higher momenta (energies). This feature is
qualitatively consistent with the results in Ref.~\cite{Dab64}. In addition, both
the $\mathcal{U}_{\mathrm{sym,1}}$ and $U_\mathrm{sym,1}$ decrease with nucleon
momentum $p$ and become negative when the nucleon momentum is larger than about $p=470$ MeV/c
(i.e., $\mathcal{E}=90$ MeV). These are in qualitative agreement with the previous
results~\cite{LiBA04,ChenLW05,LiZH06,XuC10}. Furthermore, it is interesting to see
that while the $\mathcal{U}_{\mathrm{sym,2}}$ is almost a constant, the $U_\mathrm{sym,2}$
increases with the nucleon momentum $p$ due to the transformation relation in Eq.~\eqref{b12}.
Our results indicate that the magnitude of the $U_\mathrm{sym,2}$ at $\rho_0$ can be
comparable with that of the $U_\mathrm{sym,1}$, especially at higher energies. These
features may be useful for constraining the model parameters of isospin-dependent
nuclear effective interactions.

\begin{figure}
\centering
\includegraphics[width=8.cm]{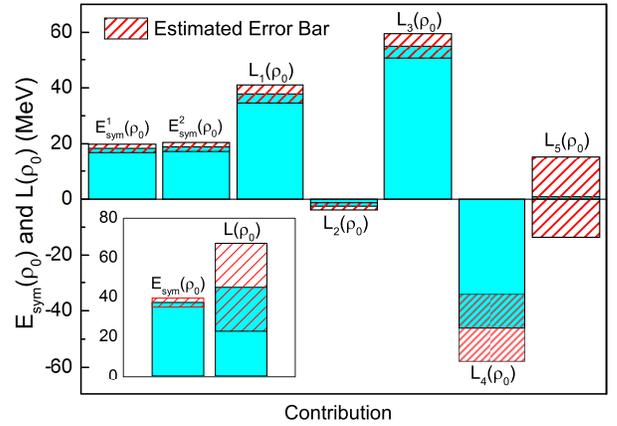}
\caption{(Color online) Contributions of each term in the decompositions
of Eq.~\eqref{a10} and Eq.~\eqref{a11} for $E_{\mathrm{sym}}(\rho_0)$ and
$L(\rho_0)$. The inset shows the results for $E_{\mathrm{sym}}(\rho_0)$ and
$L(\rho_0)$.}
\label{EsymL}
\end{figure}

From the $U_\mathrm{sym,1}$ and $U_\mathrm{sym,2}$ at $\rho_0$ together with
the empirical properties of symmetric nuclear matter at saturation density,
i.e, $\rho_0=0.16\pm0.02$ fm$^{-3}$ and $\mathcal{E}_0=-16\pm0.5$ MeV, the two
contributions to the symmetry energy are found to be
$E^1_{\mathrm{sym}}(\rho_0)=18.36\pm 1.49$ MeV and
$E^2_{\mathrm{sym}}(\rho_0)=18.88\pm 1.62$ MeV, respectively.
The five terms contributing to the slope parameter $L(\rho_0)$ are, respectively,
$L_1(\rho_0)=37.76\pm 3.23$ MeV, $L_2(\rho_0)=-2.57\pm 1.29$ MeV,
$L_3(\rho_0)=55.08\pm 4.48$ MeV, $L_4(\rho_0)=-46.11\pm 11.87$ MeV, and
$L_5(\rho_0)=0.81\pm 14.48$ MeV. These different contributions are also
shown in Fig.~\ref{EsymL} together with the $E_{\textrm{sym}}(\rho_0)$ and
$L(\rho_0)$. The resulting total symmetry energy and its density slope at
$\rho_0$ are, respectively,
\begin{align}
E_{\textrm{sym}}(\rho_0)&=37.24\pm2.26\,\textrm{MeV},\label{a32}\\
L(\rho_0)&=44.98\pm22.31\,\textrm{MeV}.
\label{a33}
\end{align}
While they are well consistent with the available constraints, the $L(\rho_0)$ is
closer to the lower end while the $E_{\textrm{sym}}(\rho_0)$ is closer to the upper
end of the range covered by previous results obtained from analyzing many other
observables using various methods (See, e.g., Refs.~\cite{LCK08,Tsa12,Lat12,ChenLW12,LiBA12}).

For the decomposition of $L(\rho_0)$, it is seen that the magnitude of $L_2(\rho_0)$
is very small, indicating that the contribution of the momentum dependence of the isoscalar
nucleon effective mass is unimportant, consistent with the theoretical calculations
in Ref.~\cite{Che12} and the assumption made in Ref.~\cite{XuC10}. On the other hand,
the $L_4(\rho_0)$ contributes a significant negative value to $L(\rho_0)$, demonstrating
the importance of the momentum dependence of the first-order symmetry potential
$U_\mathrm{sym,1}$ which is related to the isovector nucleon effective mass~\cite{XuC10}.
In addition, it is interesting to see that the magnitude of $L_5(\rho_0)$ is quite small,
consistent with the assumption made in Ref.~\cite{XuC10} where the $L_5(\rho_0)$,
i.e., the contribution of the second-order symmetry potential $U_\mathrm{sym,2}$
was neglected. The smallness of $L_5(\rho_0)$ is due to the fact that the
$U_\mathrm{sym,2}$ essentially vanishes around $\mathcal{E}=-16$ MeV as
shown in Fig.~\ref{Usym} (b). However, it should be noted that the $L_5(\rho_0)$ has a
large uncertainty, and actually it contributes the main part of the uncertainty
of $L(\rho_0)$. One can see from Fig.~\ref{EsymL} that more precise information
on the momentum dependence of $U_\mathrm{sym,1}$ and the magnitude of
$U_\mathrm{sym,2}$ at the single-nucleon energy of $\mathcal{E}_0=-16\pm0.5$ MeV
is necessary to reduce the uncertainties of the $L(\rho_0)$.

In the above evaluation of the $E_{\textrm{sym}}(\rho_0)$ and $L(\rho_0)$, the
empirical values of $\rho_0=0.16\pm0.02$ fm$^{-3}$ and $\mathcal{E}_0=-16\pm0.5$ MeV
have been used. According to the HVH theorem, the $\mathcal{E}_0$
and $\rho_0$ should be related to each other if the $U_0(\mathcal{E})$ is known.
Using $\rho_0=0.16\pm0.02$ fm$^{-3}$ and the $U_0(\mathcal{E})$ extracted in the
present work, we obtain $\mathcal{E}_0=-28.52\pm4.85$ MeV,
$E_{\textrm{sym}}(\rho_0)=40.12\pm2.84$ MeV and $L(\rho_0)=44.61\pm24.40$ MeV.
It is seen that the value of $\mathcal{E}_0=-28.52\pm4.85$ MeV significantly
deviates from the empirical value and this seems to be a common problem for
usual phenomenological optical model potentials (See, e.g.,
Ref.~\cite{Kon03,Bec69,Var91,Wep09,Han10,LiX12,Jeu76}). Fortunately, the extracted
value of $L(\rho_0)$ and its single-nucleon potential decomposition essentially
do not change although the $E_{\textrm{sym}}(\rho_0)$ increases by about $3$ MeV
compared to the results using the empirical values of $\rho_0=0.16\pm0.02$ fm$^{-3}$
and $\mathcal{E}_0=-16\pm0.5$ MeV.

\section{Summary and outlook}

In summary, using the available experimental data from neutron-nucleus scatterings,
we first obtained a new set of the global isospin dependent neutron-nucleus optical
model potential parameters which include the nuclear symmetry potential up
to the second order in isospin asymmetry for the first time. Using the analytical
relationship between the single-nucleon potentials in asymmetric nuclear matter
and the optical model potentials, we then evaluated both the first-order
and the second-order symmetry potentials $U_{\mathrm{sym,1}}$ and $U_{\mathrm{sym,2}}$
in asymmetric nuclear matter at saturation density. It is found that the strength of
the $U_\mathrm{sym,2}$ is significant compared to
that of the $U_\mathrm{sym,1}$, especially at high nucleon momentum.
Moreover, while the $U_\mathrm{sym,1}$ decreases with nucleon momentum, the
$U_\mathrm{sym,2}$ displays an opposite behavior. It will be
interesting to investigate effects of the $U_\mathrm{sym,2}$ in heavy ion collisions
induced by neutron-rich nuclei at intermediate energies. The extracted $U_{\mathrm{sym,1}}$ and
$U_{\mathrm{sym,2}}$ together with the empirical values of $\rho_0=0.16\pm0.02$
fm$^{-3}$ and $\mathcal{E}_0=-16\pm0.5$ MeV were then used to evaluate the nuclear
symmetry energy and its density slope at saturation density by applying the formulas
derived earlier based on the HVH theorem. We found that the neutron-nucleus scattering
data lead to a value of $E_{\textrm{sym}}(\rho_0)=37.24\pm2.26$ MeV and
$L(\rho_0)=44.98\pm22.31$ MeV, respectively, consistent with the results obtained
from analyzing many other observables within various models. Furthermore, our results
indicate that the contribution of the second-order symmetry potential to the
$L(\rho_0)$ is quite small, though with a large uncertainty, verifying the
assumption made in Ref.~\cite{XuC10}.

To evaluate the $E_{\textrm{sym}}(\rho_0)$ and $L(\rho_0)$ from the
single-nucleon potentials in asymmetric nuclear matter, one needs
information about the optical model potentials at negative energies, e.g.,
$\mathcal{E}_0=-16\pm0.5$ MeV. In this work, this was done by extrapolating
the optical model parameters obtained from analyzing the neutron-nucleus
scattering data in the beam energy region of $0-200$ MeV. The
extrapolation may lead to uncertainties in extracting the symmetry energy 
and its slope parameter. To be more accurate, one may use the dispersive 
optical model~\cite{Mah84,Mah87,Cha06,Cha07,LiX08} in which the real and 
imaginary parts of the optical model potential are connected with each other 
by a dispersive integration, and the optical model potentials can thus be 
extended to negative energies to address simultaneously the bound 
single-particle properties as well as elastic nucleon scattering. Using the 
dispersive optical model for elastic nucleon scattering data and the 
bound-state data of finite nuclei, one expects to obtain more reliable 
information about the optical model potentials at negative energies. It 
would be interesting to see how our present results change if the dispersive 
optical model is used. These studies are in progress and will be reported 
elsewhere.

\begin{acknowledgments}

One of us (X.H. Li) would like to thank Professors Chong-Hai Cai, Qing-Biao Shen and
Yin-Lu Han for useful discussions. This work was supported in part by the
NNSF of China under Grant Nos. 10975097, 11047157, 11135011, 11175085, 11275125,
11205083, and 11235001, the Shanghai Rising-Star Program under grant No. 11QH1401100,
the ``Shu Guang" project supported by Shanghai Municipal Education Commission and
Shanghai Education Development Foundation, the Program for Professor of Special
Appointment (Eastern Scholar) at Shanghai Institutions of Higher Learning, the
Science and Technology Commission of Shanghai Municipality (11DZ2260700), the US
National Aeronautics and Space Administration under grant NNX11AC41G issued through
the Science Mission Directorate, and the US National Science Foundation under Grant
No. PHY-1068022.

\end{acknowledgments}

\end{document}